\documentclass[aps,showkeys,showpacs,nosuperscriptaddress,twocolumn]{revtex4}

\usepackage{psfrag}
\usepackage{graphicx}
\usepackage{amsmath}
\usepackage{hyperref}
\usepackage{color}

\begin{document}

\title{If others jump to the queue front, how long I will wait?}

\author{Ma{\l}gorzata~J.~Krawczyk}
\email{gos@fatcat.ftj.agh.edu.pl}
\affiliation{
AGH University of Science and Technology, Faculty of Physics and Applied Computer Science, al. Mickiewicza 30, PL-30059 Krakow, Poland 
}

\author{Piotr~Gronek}
\email{gronek@agh.edu.pl}
\affiliation{
AGH University of Science and Technology, Faculty of Physics and Applied Computer Science, al. Mickiewicza 30, PL-30059 Krakow, Poland 
}

\author{Maria~Nawojczyk}
\email{maria@list.home.pl}
\affiliation{
AGH University of Science and Technology, Faculty of Humanities, al. Mickiewicza 30, PL-30059 Krakow, Poland 
}

\author{Krzysztof~Ku{\l}akowski}
\email{kulakowski@fis.agh.edu.pl}
\affiliation{
AGH University of Science and Technology, Faculty of Physics and Applied Computer Science, al. Mickiewicza 30, 30-059 Krakow, Poland 
}

\date{\today}

\begin{abstract}
Two models of a queue are proposed: a human queue and two lines of vehicles before a narrowing. In both models, a queuer tries to evaluate his waiting time, taking into account 
the delay caused by intruders who jump to the queue front. As the collected statistics of such events is very limited, the evaluation can give very long times. The results provide 
an example, when direct observations should be supplemented by an inference from the context.
\end{abstract}



\maketitle

\section{Introduction}
\label{S1}

In sociology, theorems are formulated rather rarely; the Thomas theorem is such an exception \cite{mert}. The theorem states: 'if men define situations as real, they are 
real in their consequences' \cite{ttt}. This reference to human mind, ubiquitous as it is in social sciences, seems to deserve more attention in sociophysical papers, where
the first principle is to keep the model simple. (For the discussion of advantages and drawbacks of this principle we refer to \cite{bru}.) Here we are going to reproduce the 
way of thinking of a rational agent who is not able to infer from a context; instead, he defines the situation as it appears from direct observations. The problem is that the 
data he can acquire in a short time are limited to one or two numbers, while a proper decision should be based on a much larger statistics. Yet, for some reasons he is willing 
to take decision at once. We ask, what can he deduce from limited data?\\

Inference of this kind, although regular in real life, is rarely considered in scientific papers, with \cite{paki} as an exception. The so-called German tank problem \cite{gua} 
bears also some resemblance to this concept. Yet, the ground is prepared by the idea of evolutionary game theory \cite{egt1,egt2,egt3}, where agents decide on the basis of an 
incomplete information. For games, the list of possible strategies is set by the definition of a given game. Here, this knowledge is substituted by a formulation of a model of the time
evolution of a given system. The task of an agent is to evaluate the model parameters.\\

As an example of a social system, here we consider a queue. This example is by no means new. Not pretending to a completeness, we provide a few examples of experimental papers. 
In a study on a long, overnight queue for football tickets in Melbourne, Leon Mann indicates that this queue is 'a miniature social system (...), formulating its own informal 
rules' \cite{mnn}. Two such rules have been identified there: $i)$ first come, 
first served, called 'a fundamental concept of queueing', and $ii)$ the right of a temporal absence, necessary in long queues. In \cite{mnt}, a cognitive bias towards unjustified 
optimism has been identified in the same Melbourne queue at persons with little chances to get the tickets. Various methods of norm executions in queues have been classified by 
Stanley Milgram and colleagues in \cite{milg}. Ellen Langer has investigated the level of 'mindlessness' in queues: as she demonstrated, people accept jumping to the queue front
when requested, even if the request is formulated in a nonsensical way \cite{lng1,lng2}.\\

Here we propose to consider two simple models of the situation where a queuer observes that somebody jumps to the queue front. The queuer tries to evaluate his expected time of waiting,
taking into account the estimated frequency of the jumping. These evaluations are to be performed on the basis of the observed data; yet these data are assumed here to be limited to
one or two observations. In both models, the queuer is able to perform some calculations or simulations; here we assume that the results of these simulations is all what
he can have. In other words, the queuer does not try to infer from the context: the physical or mental state of other queuers, publicly accessible opinions on waiting time and so on.
To be more specific, we can assume that the queuer relies on an application at his smartphone, which is able to evaluate the time of waiting in a queue in the presence of intruders.\\

In two subsequent sections, two models are presented: one of a queue of persons, and another one of two lines of vehicles before a street narrowing. For both models, the waiting 
time is found numerically, either as a function of the model parameters or at least in the form of the probability distribution. In both cases, the results indicate that the 
evaluation can give very large values of the waiting time. The last section is devoted to the discussion of an expected human reaction to the result. 

\section{A human queue}
\label{S2}

Suppose that an agent arrives to a queue for a taxi at an airport and mounts at its end. According to his expectation, the waiting time is just the product of the queue length 
$n$ and the mean time $t$ between arrivals of two subsequent vehicles. This evaluation ceases, however, its validity if the observer sees
an intruder who jumps to the queue front and takes an arriving taxi. Why he is allowed to do so? Perhaps it is because his military uniform? wonders our queuer, strange in the new 
country. Will others appear like that? How often?\\

 \begin{figure}[!hptb]
\begin{center}
\includegraphics[width=\columnwidth]{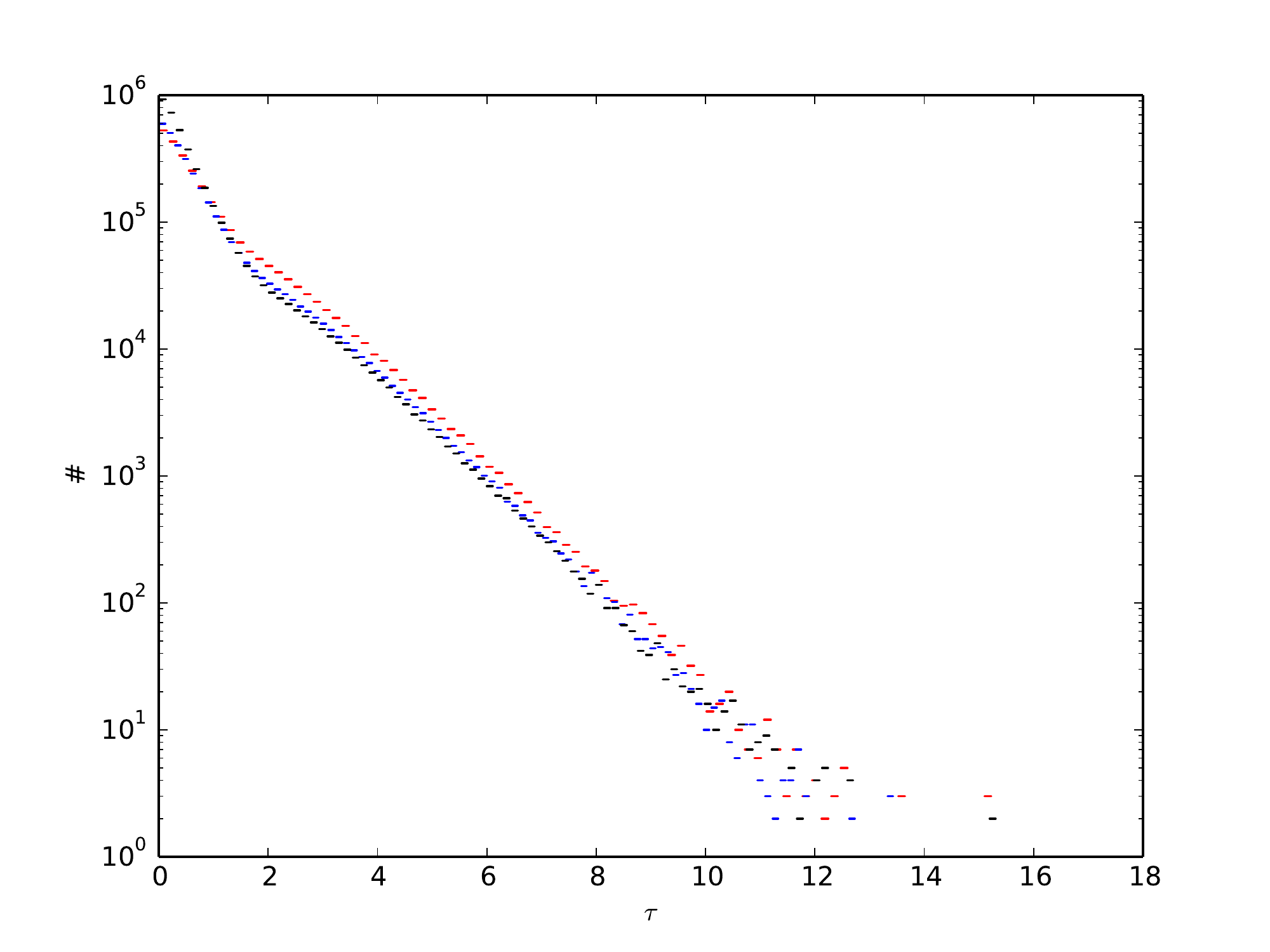}
\caption{The histogram of waiting times $\tau$ for a human queue, obtained with the algorithm in Section \ref{S2} for $a$=2,3,5. The probabilities of negative $\tau$ for these values
of $a$, obtained with the same algorithm, are 0.33, 0.25 and 0.17, respectively. The slope of the curve tail is close to -1 (note the different bases of logarithms for the axis).}
\label{fig1}
\end{center}
\end{figure}

If the mean time between intruders is $t'$, the waiting time $\tau$ can be found from

\begin{equation}
 \frac{\tau}{t}=n+\frac{\tau}{t'},
\end{equation}
where the ratio $\tau/t'$ is the expected number of intruders during the time $\tau$. Then, 

\begin{equation}
 \tau=\frac{ntt'}{t'-t}
\end{equation}
The value of the latter expression heavily depends on the observed difference $t'-t$, and it can be arbitrarily large if the observed $t'$ is close to $t$. Actually, it is even possible 
that $\tau<0$, which means that an intruder is observed earlier, than a taxi. Provided that both these times are exponentially distributed, $<t>=1/a$ and $<t'>=1$, the probability of such 
an event is

\begin{equation}
 P(t'<t)=a\int_0^\infty dt e^{-at}\int_0^t dt' e^{-t'}=\frac{1}{1+a}
\end{equation}
With this probability, the application gives an infinite waiting time. Putting this strange result aside, we calculate the probability distribution of the waiting time. 
The proposed algorithm is as follows:\\

1. Find a few random moments $t(i)$ with the exponential pdf (probability density function) with $<t(i)>=1$ ($n$-th time when somebody leaves is a sum of first $n$ moments) \\

2. Find a random moment $t'$ when the intruder appears, also with the exponential pdf with $<t'>=a$, where $a>1$\\

3. Find $t$ as the mean $t(i)$ from the events $t(i)$ with the sum smaller than $t'$\\

4. Apply Eq. (2) to calculate $\tau/n$; if the result is negative, add 1 to the 'number of negative results'. If $\tau >0$, add the result to the histogram of $(\tau)$.\\

The results on the number of negative results coincide nicely with the Eq. (3) for $a$= 2, 3 and 5. The histogram obtained for $\tau >0$ and the statistics of $10^7$ runs 
is shown in Fig. 1. As we see, the result is scale-free. This means, that very long waiting times are possible.\\

\section{Queue of two lines of vehicles}
\label{S3}

Suppose that two lines of vehicles wait before a street narrowing. To pass the narrowing, vehicles in the left line have only to reach it. On the contrary, vehicles in the 
right line have additionally to change the line, because the narrowing blocks their line. Each time when a vehicle in the left line passes the narrowing, a gap opens
in the left line between the car which moves and the car immediately behind. If a car from the right line profits the gap and changes the line, the cars in the left line
behind him do not move at all; the chance is lost till the next time. The longer the queue, the smaller the chance to move for the cars in the left line. \\

 \begin{figure}[!hptb]
\begin{center}
\includegraphics[width=\columnwidth]{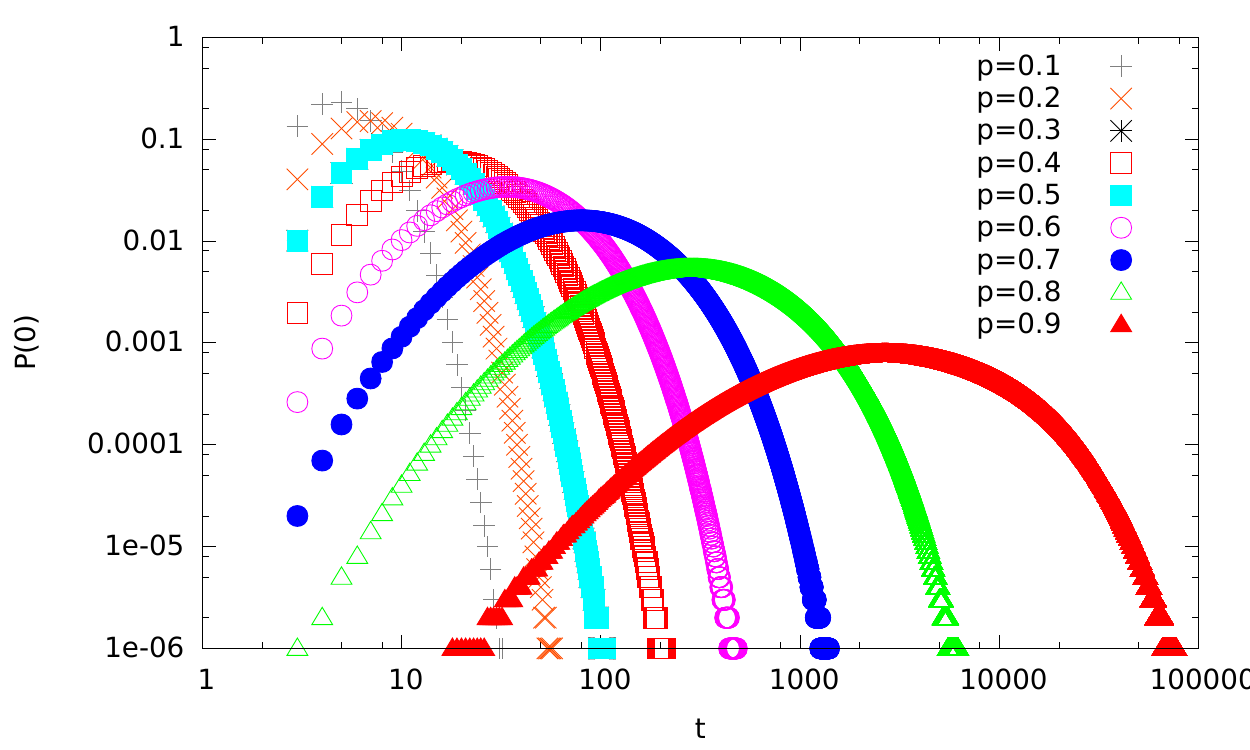}
\caption{Time dependence of $P(0)$ for the queue of vehicles. This probability, multiplied by $q$ at the next time step, is a measure of the vehicle probability flow over the narrowing. 
The parameter $p=1-q$ is the probability that a vehicle of the right line enters to a gap.}
\label{fig2}
\end{center}
\end{figure}

 \begin{figure}[!hptb]
\begin{center}
\includegraphics[width=\columnwidth]{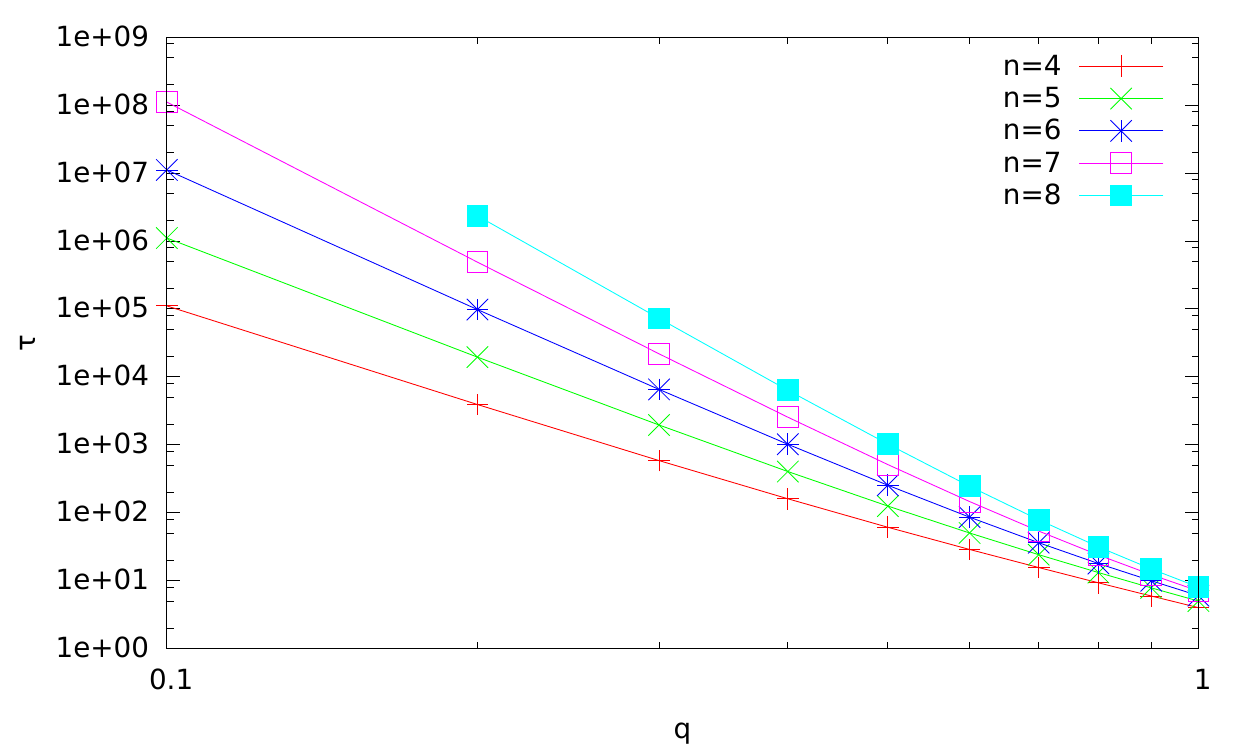}
\caption{The mean waiting time of vehicles, as dependent on the parameter $q=1-p$. For $q=1$, the waiting time $\tau=n$. However, as $p$ increases, $\tau$ can increase to astronomical 
values.}
\label{fig3}
\end{center}
\end{figure}

Consider the time-dependent probability $P_t(n)$ that at time $t$, $n$ vehicles are between a given car and the narrowing in the left line. Let us also denote by $p=1-q$ the 
probability that a car from the right line enters into a moving hole. The appropriate Master equation is 

\begin{equation}
 P_{t+1}(n)=P_t(n)(1-q)\sum_{k=0}^n q^k +P_t(n+1)q^{n+2}
\end{equation}
The first term on the r.h.s. of this equation is related to all possible events which prevent our car in the left line to move. This can happen at each of $n$ positions of the gap;
yet, once occupied, the gap disappears. The second term is related to a successful move of our vehicles from the $n+1$-th position to the $n$-th position. This needs the passive 
reaction of all $n+2$ vehicles, which can be realized in only one way. Summing up the first term, we get

\begin{equation}
 P_{t+1}(n)=P_t(n)(1-q^{n+1})+P_t(n+1)q^{n+2}
\end{equation}
The mean waiting time can be calculated as 

\begin{equation}
\tau(n,q)=q\sum_{t=0}^\infty (t+1) P_t(0)
\end{equation}
as it measures the current of probability of vehicles through the narrowing at time $t$, summarized over $t$. In Fig. 2, we show the time dependence of the probability $P(0)$ that
our vehicle appears just before the narrowing. The mean waiting time $\tau$, calculated numerically for different $n$ and $q$, is shown 
in Fig. 3. It is clear that once $q=1$, the waiting time is just $\tau=n$. Yet, for small $q$ the waiting time $\tau$ appears to grow without limits.\\

\section{Discussion}
\label{S4}

What kind of reaction for such a pessimistic result we can expect? We have no data related directly to our problem, yet some analogy can be drawn with an expected reaction
for a deterioration of conditions of a job. According to Albert Hirschman, active options are 'exit or voice', with loyalty or neglect as passive ones \cite{hrs,rus}. For
a human queue, the active options remain valid, while it might not be possible to exit from a traffic jam. The strategy 'voice' can be further differentiated to distinguish 
between an attempt to negotiate and a pure aggression. \\

Coming back to the option of inference from the context, we note that yet another option is to reject the result. One can say to himself: 'if it had looked like that, someone 
would have reacted'.  This can be chosen more likely and is more justified, because - according to our assumptions - 
the inference has been based on an incomplete information. However, other factors can play a role at least not less important. As indicated above, when we are placed at the 
end of a queue which is apparently too long to provide goods, we are prone to an unjustified optimism \cite{mnt}. This experimental result coincides with a more recent research
of attitudes of beginners in business: they are found to see only positive examples and to have an illusion of control of their life \cite{caro}. It remains not clear,
if this optimism is induced by the situation or if the beginners in business are optimistic by their nature. \\

Summarizing, we have considered two examples of model queues within the more general frames of the problem of inference from very limited data. We demonstrated that for 
both models, it may happen that the evaluated waiting time is extremely long. In our opinion, in everyday life such evaluations are usually verified on the basis of context. 
In reality both social situations provide a cognitive infrastructure in which according to Krueger \cite{kru} we are able to address personal attitudes, social norms, self-efficacy 
and collective efficacy. On the contrary, this kind of results can influence actions of AI systems, where an inference from context is not readily available.

\begin{acknowledgments}
We are grateful to Grzegorz Hara\'nczyk and Wojciech S{\l}omczy\'nski for pointing out Refs. \cite{paki,gua}. The work was partially supported by the Polish Ministry of Science and Higher Education and its grants for Scientific Research, and by the PL-Grid Infrastructure.
\end{acknowledgments}

\end{document}